\documentclass{article}

\usepackage{authblk}

\usepackage{siunitx}
\usepackage{amsmath}

\usepackage[sort,square,numbers]{natbib}

\usepackage{url}

\usepackage{graphicx}

\begin{document}

\title{\textbf{Exciton-phonon interactions in nanocavity-integrated monolayer transition metal dichalcogenides}}
\author[1]{David Rosser}
\author[2]{Taylor Fryett}
\author[2]{Albert Ryou}
\author[2]{Abhi Saxena}
\author[1,2,*]{Arka Majumdar}
\affil[1]{Department of Physics, University of Washington, Seattle, Washington 98195, USA}
\affil[2]{Department of Electrical and Computer Engineering, University of Washington, Seattle, Washington 98195, USA}
\affil[*]{Corresponding Author: arka@uw.edu}

\maketitle

\begin{abstract}
Cavity-integrated transition metal dichalcogenide excitons have recently emerged as a promising platform to study strong light-matter interactions and related cavity quantum electrodynamics phenomena. While this exciton-cavity system is typically modeled as coupled harmonic oscillators, to account for the rich solid-state environment the effect of exciton-phonon interaction needs to be incorporated. We model the system by including a phenomenological deformation potential for exciton-phonon interactions and we elucidate the experimentally measured preferential coupling of the excitonic photoluminescence to the cavity modes red-detuned with respect to the exciton resonance. Furthermore, we predict and experimentally confirm the temperature dependence of this preferential coupling. By accurately capturing the exciton-phonon interaction, our model illuminates the potential of cavity-integrated transition metal dichalcogenides for development of low-power classical and quantum technologies.
\end{abstract}

\section{Introduction}

Atomically thin van der Waals materials have recently emerged as a promising platform for engineering novel light-matter interactions \cite{xia_two-dimensional_2014, liu_van_2016, liu_van_2019}. Among various van der Waals materials, the semiconducting transition metal dichalcogenides (TMDCs) allow for the robust exploration of cavity quantum electrodynamics (cQED) in an ostensibly scalable system by their direct integration with on-chip, planar nanophotonic cavities \cite{fryett_cavity_2018, liu_van_2019}. In this hybrid system, the TMDC excitons evanescently couple to the photonic modes. Several recent studies have reported cavity integration of these materials exhibiting optically pumped lasing \cite{wu_monolayer_2015, ye_monolayer_2015, salehzadeh_optically_2015, li_room-temperature_2017}, cavity enhanced second harmonic generation \cite{fryett_silicon_2016,gan_microwatts_2018}, cavity enhanced electroluminescence \cite{liu_nanocavity_2017}, and strong coupling \cite{zhang_photonic-crystal_2018}. Beyond these demonstrations, there exist theoretical proposals utilizing TMDC excitons for quantum optical applications in single photon non-linear optics \cite{walther_giant_2018, ryou_strong_2018}. 

A necessary step for elucidating the potential applications of cavity-integrated TMDCs is an understanding of the relevant underlying physics of the exciton-cavity interaction. The prevailing description of the interaction between TMDC excitons and quantum optical cavity modes largely neglects the role of the solid state environment. However, exciton-phonon interactions are known to have a significant effect on the neutral exciton photoluminescence (PL) \cite{moody_intrinsic_2015, robert_exciton_2016, cadiz_excitonic_2017, ajayi_approaching_2017, christiansen_phonon_2017, shree_observation_2018, glazov_phonon_2019}. In other solid-state cQED systems, such as self-assembled quantum dots coupled to nanocavities, the exciton-phonon interaction is known to cause an asymmetric photoluminescent lineshape in the
form of phonon sidebands, as well as modify the cavity-coupled photoluminescence \cite{majumdar_phonon_2011, englund_resonant_2010, ates_non-resonant_2009}. In addition, after a careful review of the published literature on TMDCs coupled to whispering gallery mode resonators, we find multiple instances where the exciton's pholuminescence emission into the cavity modes appear preferentially coupled to the red-detuned side of the exciton resonance \cite{ye_monolayer_2015, salehzadeh_optically_2015, javerzac-galy_excitonic_2018, rosser_high-precision_2019}. This intriguing asymmetric coupling between the excitons and the cavity modes, which is not predicted by the simple coupled oscillator exciton-cavity theory, points to an important missing parameter in the model. 

In this paper we investigate the role of phonons in the coupling of monolayer TMDCs to nanophotonic resonators. Coupling of the TMDC neutral exciton to a cavity mode is represented as coupled oscillators within the rotating wave approximation \cite{carreno_excitation_2016, ryou_strong_2018} and a deformation potential is used to model the exciton-phonon interaction \cite{takagahara_localization_1985, singh_excitation_1994}, similar to the studies in self-assembled quantum dots coupled to nanocavities. An effective master equation is employed to describe phonon-mediated decay processes and incoherent exciton-cavity coupling \cite{hughes_phonon-mediated_2013}. Experimentally, we placed monolayer $\text{WSe}_{2}$ onto a silicon nitride ring resonator which allows for the simultaneous measurement of multiple cavity modes at different detunings. Our model exhibits preferential coupling of the exciton emission to red-detuned cavity modes, faithfully reproducing the experimental data. We further validate the theoretical model with a prediction and experimental confirmation that the asymmetry decreases with increasing temperature.

\section{Polaron Master Equation}

A homogenous distribution of TMDC excitons and a single, dispersionless cavity mode is typically formulated in terms of a coupled oscillator model $H_{XC}$ wherein the exciton and cavity coherently interact via an exciton-cavity coupling $g$ \cite{carreno_excitation_2016,ryou_strong_2018}. The resonance frequency can be measured with respect to a rotating frame at the resonant drive frequency $\omega_{L}$. The deformation potential exciton-phonon interaction $H_{XP}$ \cite{takagahara_localization_1985, singh_excitation_1994} is similar to that seen in the spin-boson model \cite{leggett_dynamics_1987, wilson-rae_quantum_2002} or for optomechanical systems \cite{rabl_photon_2011} where the exciton number operator is coupled to a bath of harmonic oscillators $b_{q}$ with frequency $\omega_{q}$ and coupling $\lambda_{q}$. Thus the coupled system is described by the Hamiltonian $H =  H_{XC} + H_{XP}$
\begin{align*}
H_{XC} &= \hbar \Delta_{XL} a^{\dagger} a + \hbar \Delta_{CL} c^{\dagger} c + \hbar  g ( a^{\dagger} c + c^{\dagger} a ) \\
H_{XP} &= \hbar a^{\dagger} a \sum_{q} \lambda_{q} ( b_{q} + b_{q}^{\dagger} ) +  \sum_{q} \hbar \omega_{q} b_{q}^{\dagger} b_{q} 	
\end{align*} where $\Delta_{XL} = \omega_{X} - \omega_{L}$ and $\Delta_{CL} = \omega_{C} - \omega_{L}$ are the detunings of the exciton and the cavity from the laser wavelength, respectively; $a$ ($c$) is the annihilation operator for the exciton (cavity) mode. In the weak excitation regime we neglect exciton saturation and any exciton-exciton interaction. Hence, we can treat both exciton and cavity operators as bosonic. 

In order to distinguish the observed neutral exciton from the effects associated with phonon bath induced fluctuations, we use the polaron transformation $P = a^{\dagger} a \sum_{q} \frac{\lambda_{q}}{\omega_{q}} (b_{q}^{\dagger} - b_{q} )$ with $H \rightarrow e^{P} H e^{-P}$ \cite{merrifield_theory_1964}, which leads to the system Hamiltonian (Supplementary Materials) \cite{leggett_dynamics_1987, mahan_many-particle_2000,rabl_photon_2011, roy_polaron_2012}
\begin{equation}\label{ham}
H'_{S} = \hbar (\Delta_{XL} - \Delta_P) a^{\dagger} a - \hbar \Delta_{P} a^{\dagger} a^{\dagger} a a + \hbar \Delta_{CL} c^{\dagger} c + \hbar \langle B \rangle g (\sigma^{+} a + a^{\dagger} \sigma^{-})
\end{equation} The exciton resonance $\Delta_{xL}' = \Delta_{xL} - \Delta_{P}$ is renormalized by a polaron shift $\Delta_{P} = \sum_{q} \frac{\lambda_{q}^{2}}{\omega_{q}}$, which is analogous to a Lamb shift \cite{roy-choudhury_quantum_2015}. When the harmonic oscillator bath is written in terms of the phonon displacement operator $B_{\pm} = \exp \left [ \pm \sum_{q} \frac{\lambda_{q}}{\omega_{q}} (b_{q} - b_{q}^{\dagger} ) \right ] $ the exciton-cavity coupling is modified from the bare value by the average phonon displacement $\langle B \rangle$
\begin{equation*}
\langle B \rangle = \exp \left [ -\frac{1}{2} \sum_{q} \left ( \frac{\lambda_{q}}{\omega_{q}} \right )^{2} (2 \bar{n}_{q} + 1 ) \right ]
\end{equation*}
where $\bar{n}_{q} = [ e^{\beta \hbar \omega_{q}} - 1 ]^{-1}$ is the mean phonon number with bath temperature $T = 1/{k_{B} \beta}$ \cite{roy_influence_2011}. As the temperature increases, the average phonon number in each mode increases, which decreases the exciton-cavity interaction.

\begin{figure}
	\centering
	\includegraphics[width=1.0\linewidth]{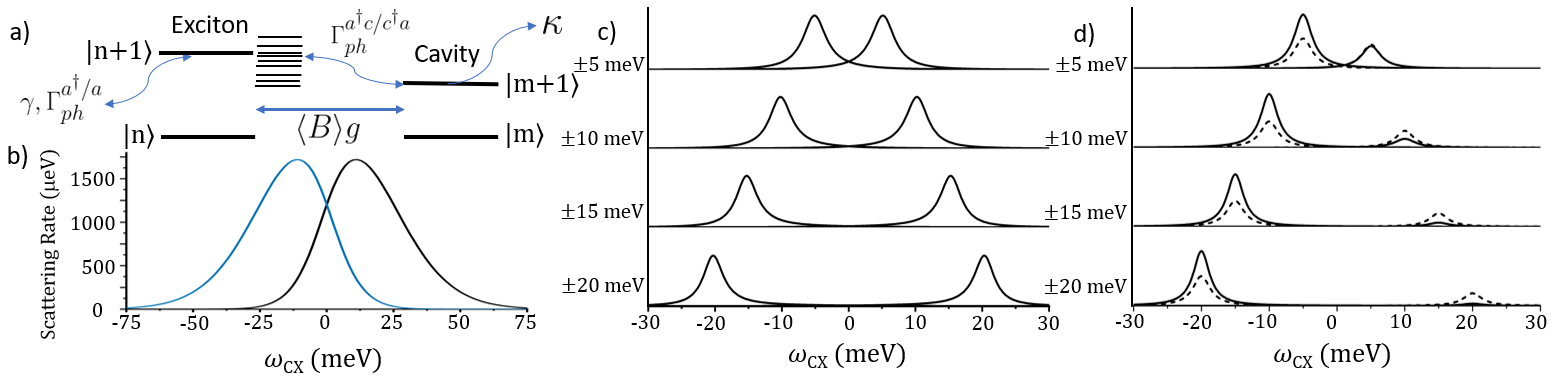}
	\caption{Theoretical modeling with the Hamiltonian described in the text ($T = 80$ \si{\kelvin}, $\gamma = 48.4$ \si{\milli\electronvolt}, $\alpha_{p} = 0.018$ \si{\pico \second}$^2$, $\omega_{b} = 6.7$ \si{\milli \electronvolt}, $\kappa = 2.85$ \si{\milli\electronvolt}, $g = 4$ \si{\milli\electronvolt}) a) Level diagram with phonon-mediated scattering. b) Asymmetric phonon-mediated exciton-cavity coupling rates. The blue line gives the phonon-mediated incoherent emission into the cavity. Note that the peak is not centered at zero detuning. c) Detuning dependent ($\Delta_{CX} = \pm 5, \pm 10, \pm 15, \pm 20$ \si {\milli \electronvolt}) cavity emission without phonons. d) Detuning dependent cavity emission with phonons at 80 \si{\kelvin} (solid line) and 320 \si{\kelvin} (dashed line). Note that for the $\Delta_{CX} = \pm 5$ \si {\milli \electronvolt} the dashed and solid line are on top of each other for the blue detuned case.}
	\label{fig:theory}
\end{figure}

We employ an effective master equation $\frac{\partial \rho}{\partial t} = \frac{1}{i \hbar} [H',\rho] + \frac{\kappa}{2}\mathcal{L}[c] + \frac{\gamma}{2}\mathcal{L}[a] + \frac{\Gamma^{a^{\dagger}c}_{ph}}{2}\mathcal{L}[a^{\dagger} c] + \frac{\Gamma^{c^{\dagger}a}_{ph}}{2}\mathcal{L}[c^{\dagger} a]$ \cite{hughes_phonon-mediated_2013} to model the incoherent exciton-cavity feeding. Figure \ref{fig:theory}a illustrates the energy-level diagram of the exciton and cavity system. The dissipator $\mathcal{L}[\xi] = \xi \rho \xi^{\dagger} - \frac{1}{2} \xi^{\dagger} \xi \rho - \frac{1}{2} \rho \xi^{\dagger} \xi$ with Lindblad operators $\xi$ describes the cavity decay rate ($\kappa$), exciton decay rate ($\gamma$), and the incoherent phonon-mediated exciton-cavity scattering ($\Gamma^{a^{\dagger}c}_{ph}$, $\Gamma^{c^{\dagger}a}_{ph}$).

The phonon-mediated exciton-cavity scattering (Fig. \ref{fig:theory}b) with cavity-exciton detuning $\Delta_{CX} = \omega_{C} - \omega_{X}$ is given by
\begin{equation}\label{scattering}
\Gamma^{a^{\dagger}c/c^{\dagger}a}_{ph} = 2 \langle B \rangle^{2} g^{2} \mathrm{Re}\left [ \int_{0}^{\infty} d\tau e^{\pm \Delta_{CX} \tau} (e^{\phi(\tau)}-1) \right ]
\end{equation} with the phonon correlation function \cite{leggett_dynamics_1987,mccutcheon_quantum_2010} 
\begin{equation*}
\phi(\tau) = \int_{0}^{\infty} d\omega \frac{J(\omega)}{\omega^{2}} \left [ \coth\left (\frac{\hbar \omega}{2 k_{B} T} \right ) \cos(\omega \tau) - i \sin(\omega \tau) \right ]
\end{equation*}We assume a Gaussian localization of the exciton confined to the monolayer TMDC due to substrate inhomogeneities for a qualitative super-ohmic spectral density $J(\omega) = \alpha_{p} \omega^{3} \exp(-\omega^{2}/2\omega_{b}^{2})$ \cite{leggett_dynamics_1987, takagahara_localization_1985} with $\alpha_{p}$ and $\omega_{b}$ serving as the exciton-phonon coupling strength and cutoff frequency, respectively. This phonon spectral function is identical to that used in quantum dot studies of phonon interactions \cite{calarco_spin-based_2003, nazir_photon_2008, roy_polaron_2012}.

Without phonon-mediated scattering the peak cavity intensity occurs at zero detuning ($\omega_{C} = \omega_{X}$) and the cavity-coupled PL is symmetric with respect to the exciton PL emission peak (Fig. \ref{fig:theory}c). The additional scattering from phonon processes of the exciton into the cavity mode dominates when the cavity is red-detuned with respect to the exciton (Fig. \ref{fig:theory}d). Physically, we expect down-conversion of an exciton into a phonon and cavity photon as an example of a Stokes process. The opposite up-conversion amounts to optical refrigeration \cite{jones_excitonic_2016}. Including phonon-mediated scattering demonstrates the peak cavity intensity is red-detuned with respect to the exciton PL emission peak. Furthermore, our model predicts that at the same detuning, the relative intensity between the red-detuned and blue-detuned cavity-coupled photoluminescence decreases for increasing temperature (Fig. \ref{fig:theory}d, dashed line) \cite{hughes_phonon-mediated_2013}.

\section{Exciton-Cavity Photoluminescence Spectra}

\begin{figure}
	\centering
	\includegraphics[width=1.0\linewidth]{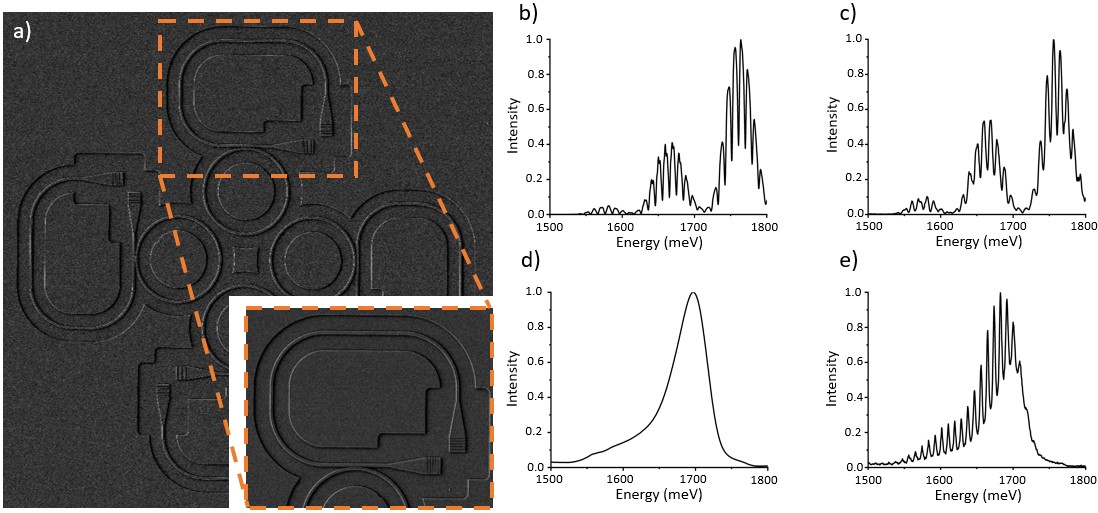}
	\caption{a) SEM of four 5 \si{\micro \meter} (radius) uncoupled SiN ring resonators. Inset: SEM of the coupled ring/waveguide and grating couplers. The grating couplers are used to input light and collect transmitted light. b) Transmission spectrum of the SiN ring resonator before integration of monolayer $\text{WSe}_{2}$. c) Transmission spectrum of the SiN ring resonator after integration of monolayer $\text{WSe}_{2}$. d) PL of monolayer $\text{WSe}_{2}$. e) Cavity-coupled PL of monolayer $\text{WSe}_{2}$.}
	\label{fig:initial}
\end{figure}

To validate our quantum optical model, we performed experiments with a ring resonator integrated with a monolayer of $\text{WSe}_{2}$. A ring resonator can support multiple cavity modes separated by the free spectral range, and thus provides an ideal platform for studying the coupling of the photoluminescence to cavity modes with different detunings from the exciton. The ring resonator is fabricated using a 220 \si{\nano \meter} thick silicon nitride (SiN) membrane grown via low pressure chemical vapor deposition (LPCVD) on 4 \si{\micro \meter} of thermal oxide on silicon. We spun roughly 400 \si{\nano \meter} of Zeon ZEP520A, which was coated with a thin layer of Pt/Au that served as a charging layer. The resist was patterned using a JEOL JBX6300FX electron-beam lithography system with an accelerating voltage of 100 \si{\kilo \volt}. The pattern was transferred to the SiN using a reactive ion etch (RIE) in $\text{CHF}_{3} \text{/O}_{2}$ chemistry. The mechanically exfoliated $\text{WSe}_2$ was then transferred onto the SiN ring resonator (Fig. \ref{fig:initial}a) using a modified dry transfer method to eliminate bulk material contamination \cite{rosser_high-precision_2019} which would otherwise quench the optical properties of the waveguide structures.

The transmission spectrum of the SiN ring resonator is measured by exciting a grating coupler with a supercontinuum laser (Fianium WhiteLase Micro) and collecting from the other grating coupler (Fig. \ref{fig:initial}a, inset). An initial transmission measurement of the ring resonator before monolayer TMDC transfer yields the bare cavity linewidth of $\kappa = 2.85$ \si{\milli\electronvolt} (Fig. \ref{fig:initial}b). The dips in the transmission correspond to the resonance in the ring resonators. The separation between the modes corresponds to the free spectral range ($\text{FSR}  = \frac{c}{2 \pi n_{eff} R} \approx 4.8$ \si{\tera\hertz}) of the ring resonator, which matches the FSR expected from the ring radius ($R = 5$ \si{\micro\meter}) and effective index of refraction of the SiN waveguide ($n_{eff} \approx 2$). The envelope modulation of the spectrum is due to the frequency-dependent coupling efficiency of the grating couplers (Figs. \ref{fig:initial}b and \ref{fig:initial}c). The angular dependence of the grating coupler does not affect the cavity-coupled PL measurement due to the large numerical aperture of our objective lens. There exists a relative amplitude change between the envelope modulation function in the observed transmission spectrum due to the angular dependence of the grating couplers. As the measurement is done before and after the transfer, which requires removing the sample from the optical setup, the angular alignment of the confocal microscope objective to the grating coupler will be slightly different \cite{chrostowski_silicon_2015}. The transmission spectrum of the ring resonator after material transfer demonstrates the monolayer does not significantly affect the cavity modes (Fig. \ref{fig:initial}c). It is important to point out that with the exciton at approximately 1700 \si{\milli \electronvolt}, the small linewidth increase seen in the transmission spectrum equally affects cavity modes both red and blue-detuned with respect to the exciton resonance.

Photoluminescence was first measured to confirm the existence of the monolayer after material transfer because 2D materials exhibit poor optical contrast on the SiN substrate (Fig. \ref{fig:initial}d). The strong excitonic peak of the $\text{WSe}_2$ monolayer integrated onto the SiN ring resonator establishes the presence of the vdW material on the waveguide \cite{chernikov_exciton_2014}. The primary peak is attributed to neutral exciton emission. The secondary sidebands could be due to defects or trion emission \cite{chow_defect-induced_2015, sidler_fermi_2017}. PL is measured by exciting the monolayer with a HeNe laser (40 \si{\micro \watt} at 633 \si{\nano \meter}). By fitting the measured PL at 80 \si{\kelvin}, the material dependent parameters for the phonon spectral function can be calculated, independent of the cavity coupling (Supplementary Materials). We found an exciton linewidth $\gamma = 48.4$ \si{\milli\electronvolt}, an exciton-phonon coupling $\alpha_{p}= 0.018$ \si{\pico \second}$^2$, and cutoff frequency $\omega_{b} = 6.7$ \si{\milli \electronvolt}. These extracted parameters are consistent with values estimated from bulk material measurements (Supplementary Materials). The polaron shift of the exciton energy is then calculated to be $\hbar \Delta_{P} = \hbar \int_{0}^{\infty} d\omega J(\omega) / \omega = \hbar \sqrt{\frac{\pi}{2}} \alpha_{p} \omega_{b}^{3} = 24$ \si{\milli \electronvolt}, which we incorporate into the modified exciton resonance $\Delta_{xL}' = \Delta_{xL} - \Delta_{P}$.

\begin{figure}
	\centering
	\includegraphics[width=1.0\linewidth]{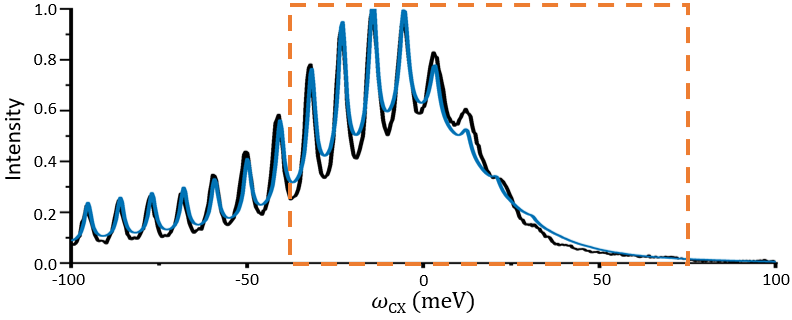}
	\caption{Measured cavity-coupled PL (black) and simulated cavity-coupled PL (blue) at 80 \si{\kelvin}. Theoretical model fit to the windowed region of data with the Hamiltonian described in the text ($T = 80$ \si{\kelvin}, $\gamma = 48.4$ \si{\milli\electronvolt}, $\alpha_{p} = 0.018$ \si{\pico \second}$^2$, $\omega_{b} = 6.7$ \si{\milli \electronvolt}, $\kappa = 2.85$ \si{\milli\electronvolt}, $g = 4$ \si{\milli\electronvolt}). }
	\label{fig:fit}
\end{figure}

Cavity-coupled PL is measured by directly exciting the monolayer $\text{WSe}_2$ from the top and collecting the resulting emission from a grating coupler using a pinhole in the image plane of a free-space confocal microscope. Cavity-coupled PL exhibits asymmetric emission into the cavity modes where there is greater intensity in the cavities red-detuned with respect to the exciton (Fig. \ref{fig:initial}e). The coherent exciton-cavity coupling $\hbar g$ can be extracted by considering the linear superposition of all cavity resonances for the ring resonator and including a contribution from background PL that is difficult to completely remove due to the proximity of the grating coupler and laser excitation of the monolayer $\text{WSe}_2$. The exciton-cavity coupling accounting for the average phonon displacement is found to be $\hbar g \approx 4-6 \ \si{\milli \electronvolt}$ (Fig. \ref{fig:fit}) by a brute force search minimizing the least squares error between the simulated and observed data over a windowed region of the cavity-coupled PL spectrum. The far red-detuned data attributed to defect and trion emission was accounted for by a convolution of the PL and Lorentzian cavity modes (Supplementary Materials). In this experiment only $\sim 1/4$ of the SiN ring resonator was covered with monolayer $\text{WSe}_2$. A full coverage of monolayer $\text{WSe}_2$ on the SiN ring resonator gives $\hbar g \approx 8-12 \ \si{\milli \electronvolt}$ as an estimated coherent interaction of the exciton and cavity mode due to the $g \propto \sqrt{N}$ scaling of the light-matter interaction in the collective excitation basis and assuming the number of available exciton states is proportional to the area of monolayer material on the cavity. Our extracted $g$ value is consistent with the light-matter interaction $g \approx 10 - 14$ \si{\milli \electronvolt} found in strong-coupling experiments with van der Waals materials integrated on photonic crystal cavities with comparable length of the cavity \cite{dufferwiel_excitonpolaritons_2015, zhang_photonic-crystal_2018, kravtsov_nonlinear_2019}. We note that for the ring resonator, the length of the cavity that goes into calculation of the $g$ is the thickness of the slab ($\sim 220$ nm).

\begin{figure}
	\centering
	\includegraphics[width=1.0\linewidth]{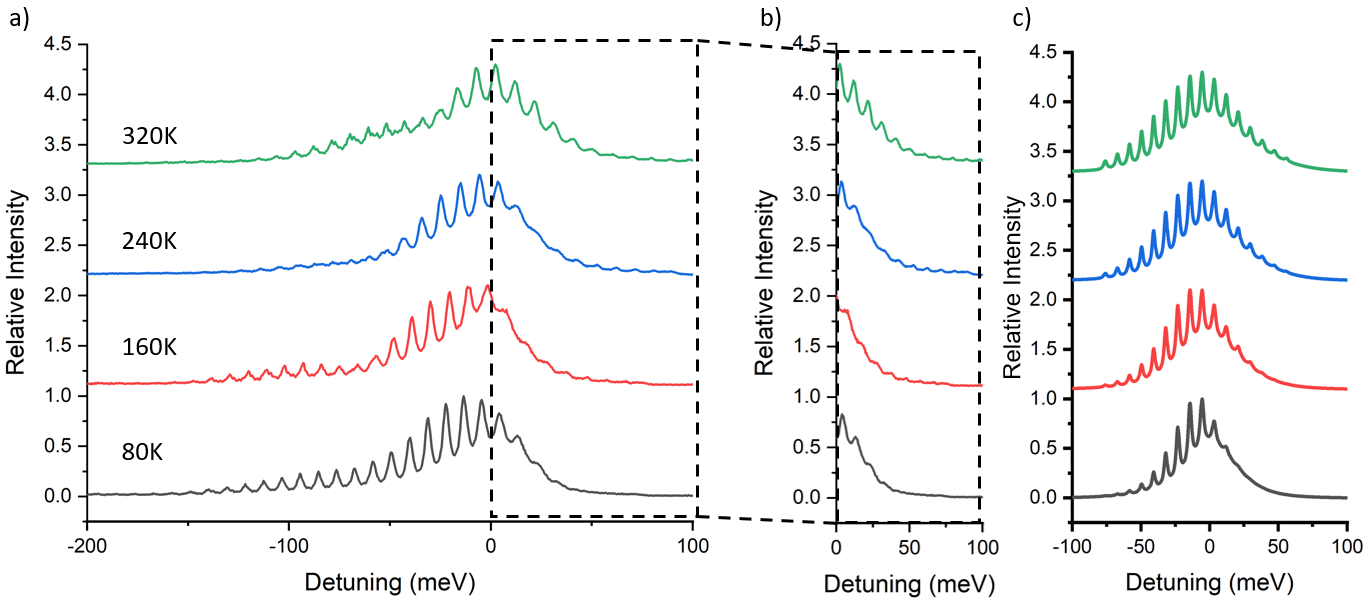}
	\caption{a) Temperature dependence (80 \si{\kelvin} to 320 \si{\kelvin}) of the asymmetric cavity-coupled PL. b) Zoomed-in to show the temperature dependence of the asymmetric cavity-coupled PL for cavities blue-detuned with respect to the exciton. c) Simulated temperature dependence of cavity-coupled PL without trion and defect emission. All free parameters are held fixed except the measured cryostat temperature ($T = 80$ \si{\kelvin}, 160 \si{\kelvin}, 240 \si{\kelvin}, 320 \si{\kelvin}, $\gamma = 48.4$ \si{\milli\electronvolt}, $\alpha_{p} = 0.018$ \si{\pico \second}$^2$, $\omega_{b} = 6.7$ \si{\milli \electronvolt}, $\kappa = 2.85$ \si{\milli\electronvolt}, $g = 4$ \si{\milli\electronvolt}).}
	\label{fig:temperature}
\end{figure}

To further confirm the theoretical model, we measure the temperature-dependent variation in the asymmetric coupling in the range 80 \si{\kelvin} - 320 \si{\kelvin}. Using liquid nitrogen in a continuous flow cryostat (Janis ST-500) we can tune the energy of the exciton in the monolayer $\text{WSe}_2$ from 1650 \si{\milli\electronvolt} - 1700 \si{\milli\electronvolt} with the consequent changes in linewidth. As the cryostat temperature is increased we see cavity-coupled PL extending to further blue-detuned cavities with respect to the exciton energy (Fig. \ref{fig:temperature}a) where the spectra are shifted by the exciton center frequency. In particular, the maximum detuning with visible cavity modes increases with increasing temperature (Fig. \ref{fig:temperature}b).  We find the model Hamiltonian parameters extracted from the PL and cavity-coupled PL qualitatively reproduce the spectrum at elevated temperatures (Fig. \ref{fig:temperature}c) where the only modified simulation parameter is the measured temperature of the cryostat. Reduced asymmetry in cavity-coupled PL at elevated temperatures is due to the reduced asymmetry of the phonon-mediated exciton-cavity coupling rates with respect to the neutral exciton resonance.

\section{Conclusion}

We have explored exciton-phonon interactions in cavity-integrated TMDCs and demonstrated that a phenomenological deformation potential has significant value in explaining asymmetric cavity emission \cite{ye_monolayer_2015, salehzadeh_optically_2015, javerzac-galy_excitonic_2018, rosser_high-precision_2019}. Reflecting on the effective system Hamiltonian (Eq. \ref{ham}), the polaron shift $\Delta_{P} = 24$ \si{\milli \electronvolt} of the exciton energy is comparable to the $\Delta_{P} = 29$ \si{\milli \electronvolt} found via the excitonic Bloch equations \cite{christiansen_phonon_2017}. Temperature dependence of the exciton-cavity coupling has been previously observed in strong-coupling experiments with TMDC excitons \cite{liu_control_2017, zhang_photonic-crystal_2018, kravtsov_nonlinear_2019}, although a rigorous model explaining this behavior was not reported.  We attribute this modification of the bare value to the average phonon displacement $g \rightarrow \langle B \rangle g$. A consequence of the exciton-cavity incoherent scattering (Eq. \ref{scattering}) is an efficient means for exciton population inversion which could potentially explain observations of lasing in cavity-integrated monolayer materials \cite{hughes_phonon-mediated_2013, wu_monolayer_2015, ye_monolayer_2015, salehzadeh_optically_2015, li_room-temperature_2017, lohof_prospects_2019}. In the interest of a low-power optical non-linearity, polaron-polaron scattering in the effective system Hamiltonain ($\Delta_{P} a^{\dagger} a^{\dagger} a a$) provides an interesting opportunity which could lead to non-classical light generation \cite{rabl_photon_2011}. The calculated polaron shift is two orders of magnitude larger than expected for exciton-exciton scattering due to a lateral confining potential \cite{ryou_strong_2018}. A full understanding of the many-body interactions in nanocavity-integrated monolayer transition metal dichalcogenides is a necessary prerequisite to assessing the potential of this system for future classical and quantum technologies.

\section{Acknowledgments}
The research was supported by NSF-1845009, NSF-ECCS-1708579 and AFOSR grant FA9550-
17-C-0017. Part of this work was conducted at the Washington Nanofabrication Facility / Molecular Analysis Facility, a National Nanotechnology Coordinated Infrastructure (NNCI) site at the University of Washington, which is supported in part by funds from the National Science Foundation (awards NNCI-1542101, 1337840 and 0335765), the National Institutes of Health, the Molecular Engineering \& Sciences Institute, the Clean Energy Institute, the Washington Research Foundation, the M. J. Murdock Charitable Trust, Altatech, ClassOne Technology, GCE Market, Google and SPTS. A.R. acknowledges
support from the IC Postdoctoral Research Fellowship.


\clearpage
\begin{center}
	\textbf{\large Supplementary Materials: Exciton-phonon interactions in nanocavity-integrated monolayer transition metal dichalcogenides}
\end{center}
\setcounter{equation}{0}
\setcounter{figure}{0}
\setcounter{table}{0}
\setcounter{page}{1}
\setcounter{section}{0}
\makeatletter
\renewcommand{\theequation}{S\arabic{equation}}
\renewcommand{\thefigure}{S\arabic{figure}}

\section{Polaron Transformation}

The microscopic Hamiltonian description of our system begins with a coupled oscillator model $H_{XC}$ wherein the exciton $\Delta_{XL} = \omega_{X} - \omega_{L}$ and cavity $\Delta_{CL} = \omega_{C} - \omega_{L}$ resonances, measured with respect to the laser frequency $\omega_{L}$, interact via an exciton-cavity coupling $g$ \cite{carreno_excitation_2016,ryou_strong_2018}. The deformation potential exciton-phonon interaction ($H_{XP}$) \cite{takagahara_localization_1985,singh_excitation_1994} is a simplified model to account for effects of the solid state environment. Although we are interested in incoherent excitation for this experiment we will later argue for a coherent driving term for the exciton $\eta_{X}$ in photoluminescence. The total Hamiltonian is $H =  H_{XC} + H_{D} + H_{XP}$ with components
\begin{align}
H_{XC} &= \hbar \Delta_{XL} a^{\dagger} a + \hbar \Delta_{CL} c^{\dagger} c + \hbar  g ( a^{\dagger} c + c^{\dagger} a ) \\
H_{D} &= \hbar \eta_{X}(a + a^{\dagger}) \\
H_{XP} &= \hbar a^{\dagger} a \sum_{q} \lambda_{q} ( b_{q} + b_{q}^{\dagger} ) +  \sum_{q} \hbar \omega_{q} b_{q}^{\dagger} b_{q} 	
\end{align} The polaron transformation $P = a^{\dagger} a \sum_{q} \frac{\lambda_{q}}{\omega_{q}} (b_{q}^{\dagger} - b_{q} )$ with $H \rightarrow e^{P} H e^{-P}$ \cite{merrifield_theory_1964,wilson-rae_quantum_2002} using the identity $e^{X} Y e^{-X} = Y + [X,Y]+ \frac{1}{2!}[X,[X,Y]]+\frac{1}{3!}[X,[X,[X,Y]]]+\cdot\cdot\cdot$ and $a$, $b$, and $c$ are all bosonic operators (e.g. $[a,a^{\dagger}]=1$) leads to
\begin{align*}
\Delta_{XL} a^{\dagger}a &\rightarrow \Delta_{XL} a^{\dagger}a \\
\Delta_{CL} c^{\dagger}c &\rightarrow \Delta_{CL} c^{\dagger}c \\ 
a^{\dagger} a \sum_{q} \lambda_{q} ( b_{q} + b_{q}^{\dagger} ) &\rightarrow a^{\dagger} a \sum_{q} \lambda_{q} ( b_{q} + b_{q}^{\dagger} ) - 2 (a^{\dagger} a)^{2} \sum_{q} \frac{\lambda_{q}^{2}}{\omega_{q}} \\
& =a^{\dagger} a \sum_{q} \lambda_{q} ( b_{q} + b_{q}^{\dagger} ) - 2 (a^{\dagger} a)^{2} \Delta_{P} \\
\sum_{q} \omega_{q} b_{q}^{\dagger} b_{q} &\rightarrow \sum_{q} \omega_{q} b_{q}^{\dagger} b_{q} - a^{\dagger} a \sum_{q} \lambda_{q} ( b_{q} + b_{q}^{\dagger} ) + (a^{\dagger} a)^{2} \sum_{q} \frac{\lambda_{q}^{2}}{\omega_{q}} \\
& = \sum_{q} \omega_{q} b_{q}^{\dagger} b_{q} - a^{\dagger} a \sum_{q} \lambda_{q} ( b_{q} + b_{q}^{\dagger} ) + (a^{\dagger} a)^{2} \Delta_{P}
\end{align*}
\begin{align*}
g ( a^{\dagger} c + c^{\dagger} a ) + \eta_{X} (a + a^{\dagger}) &\rightarrow g ( a^{\dagger} c + c^{\dagger} a ) + \eta_{X} (a + a^{\dagger}) \\
& + [\sum_{q} \frac{\lambda_{q}}{\omega_{q}}(b_{q}^{\dagger} - b_{q})][g ( a^{\dagger} c - c^{\dagger} a ) + \eta_{X} (a - a^{\dagger})] \\
& + [\sum_{q} \frac{\lambda_{q}}{\omega_{q}}(b_{q}^{\dagger} - b_{q})]^{2}[g ( a^{\dagger} c + c^{\dagger} a ) + \eta_{X} (a + a^{\dagger})] \\
& + [\sum_{q} \frac{\lambda_{q}}{\omega_{q}}(b_{q}^{\dagger} - b_{q})]^{3}[g ( a^{\dagger} c - c^{\dagger} a ) + \eta_{X} (a - a^{\dagger})] + \cdot \cdot \cdot \\
& = \exp[\sum_{q}\frac{\lambda_{q}}{\omega_{q}}(b_{q}^{\dagger} - b_{q})](g a^{\dagger} c + \eta_{X} a) + \exp[-\sum_{q}\frac{\lambda_{q}}{\omega_{q}}(b_{q}^{\dagger} - b_{q})](g c^{\dagger} a + \eta_{X} a^{\dagger}) \\
& = g(\underbrace{B_{+}a^{\dagger}c}_{1} + \underbrace{B_{-}c^{\dagger}a}_{2}) + \eta_{X} (\underbrace{B_{+}a^{\dagger}}_{3} + \underbrace{B_{-} a}_{4}) \\
& = \frac{1}{2}[\underbrace{g B_{+} c^{\dagger} a}_{I} + \underbrace{g B_{+} a^{\dagger} c}_{1} + \underbrace{\eta_{X} B_{+} a}_{III} + \underbrace{\eta_{X} B_{+} a^{\dagger}}_{3} \\
& + \underbrace{g B_{-} c^{\dagger} a}_{2} + \underbrace{g B_{-} a^{\dagger} c}_{II} + \underbrace{\eta_{X} B_{-} a}_{4} + \underbrace{\eta_{X} B_{-} a^{\dagger}}_{IV} \\
& + \underbrace{g B_{+} a^{\dagger} c}_{1} - \underbrace{g B_{+} c^{\dagger} a}_{I} + \underbrace{\eta_{X} B_{+} a^{\dagger}}_{3} - \underbrace{\eta_{X} B_{+} a}_{III} \\
& - \underbrace{g B_{-} a^{\dagger} c}_{II} + \underbrace{g B_{-} c^{\dagger} a}_{2} - \underbrace{\eta_{X} B_{-} a^{\dagger}}_{IV} + \underbrace{\eta_{X} B_{-} a}_{4} ] \\
& = \frac{1}{2}(B_{+} + B_{-}) [g(c^{\dagger} a + a^{\dagger} c) + \eta_{X}(a + a^{\dagger})] \\
& + \frac{1}{2}(B_{+} - B_{-}) [g(a^{\dagger} c - c^{\dagger} a) + \eta_{X}(a^{\dagger} - a)] \\
& = \frac{1}{2}(B_{+} + B_{-}) X_{g} + \frac{1}{2 i}(B_{+} - B_{-}) X_{u} \\
& = \frac{1}{2}(B_{+} + B_{-} - 2 \langle B \rangle ) X_{g} + \langle B \rangle X_{g} + \frac{1}{2i}(B_{+} - B_{-}) X_{u} \\
& = \zeta_{g} X_{g} + \zeta_{u} X_{u} + \langle B \rangle X_{g}
\end{align*} We define the polaron shift $\Delta_{P} = \sum_{q} \frac{\lambda_{q}^{2}}{\omega_{q}} = \int_{0}^{\infty} d\omega J(\omega) / \omega$ and the exciton-cavity coupling terms
\begin{align*}
X_{g} &= \hbar g (c^{\dagger} a + a^{\dagger} c) + \hbar \eta_{X}(a + a^{\dagger}) \\
X_{u} &= i \hbar g (a^{\dagger} c - c^{\dagger} a) + i \hbar \eta_{X}(a^{\dagger} - a)
\end{align*} $J(\omega)$ is the phonon spectral function discussed in the main text. The bath displacement operators $B_{\pm} = \exp \left [ \pm \sum_{q} \frac{\lambda_{q}}{\omega_{q}} (b_{q} - b_{q}^{\dagger} ) \right ]$ are included in the exciton-cavity prefactors
\begin{align*}
\zeta_{g} &= \frac{1}{2}(B_{+}+B_{-}-2\langle B \rangle) \\
\zeta_{u} &= \frac{1}{2 i}(B_{+}-B_{-})
\end{align*} where $\langle B \rangle=\langle B_{+} \rangle=\langle B_{-} \rangle$.
The resulting system, bath, and interaction Hamiltonian give
\begin{align}
H'_{S} &= \hbar \Delta_{XL} a^{\dagger} a - \hbar \Delta_{P} (a^{\dagger} a)^{2} + \hbar \Delta_{CL} c^{\dagger} c + \langle B \rangle X_{g} \\
&= \hbar (\Delta_{XL} - \Delta_P) a^{\dagger} a - \hbar \Delta_{P} a^{\dagger} a^{\dagger} a a + \hbar \Delta_{CL} c^{\dagger} c + \langle B \rangle X_{g} \\
H'_{B} &= \sum_{q} \hbar \omega_{q} b_{q}^{\dagger} b_{q} \\
H'_{I} &= X_{g}\zeta_{g} + X_{u}\zeta_{u}
\end{align} 

\section{Effective Phonon Master Equation}

We use the Markovian time-convolutionless (TCL) master equation to approximate the temporal dynamics of our system \cite{wilson-rae_quantum_2002,mccutcheon_quantum_2010,roy_phonon-dressed_2011,roy_influence_2011,roy_polaron_2012,manson_polaron_2016}.
\begin{equation}
\frac{\partial \rho(t)}{\partial t} = \frac{1}{i \hbar} [H'_{S}, \rho(t)] + \mathcal{L}(\rho) - \frac{1}{\hbar^{2}}\int_{0}^{\infty} d\tau \sum_{m=g,u} \{G_{m}(\tau)[\hat{X}_{m},e^{-i H'_{S} \tau / \hbar}\hat{X}_{m} e^{i H'_{S} \tau / \hbar}\rho(t)] + H.c.\}
\end{equation} $G_{m}(t)$ are the polaron Green functions ($m = g, u$)
\begin{align*}
G_{g}(t) &= \langle B \rangle^{2} {\cosh[\phi(t)]-1} \\
G_{u}(t) &= \langle B \rangle^{2} \sinh[\phi(t)]
\end{align*} and $\phi(\tau) = \int_{0}^{\infty} d\omega \frac{J(\omega)}{\omega^{2}} \left [ \coth\left (\frac{\hbar \omega}{2 k_{B} T} \right ) \cos(\omega \tau) - i \sin(\omega \tau) \right ]$ is the phonon correlation function. The average phonon displacement can be calculated as $\langle B \rangle = \exp(-\phi(0) / 2)$ \cite{manson_polaron_2016}.

To approximate the integral in the master equation we assume the cavity-exciton detuning ($\Delta_{CX}$) is large compared to $g$ and weak excitation such that $\langle a^{\dagger} a \rangle << 1$. We can then say
\begin{equation}
e^{-i H'_{S} \tau / \hbar}\hat{X}_{m} e^{i H'_{S} \tau / \hbar} \simeq e^{-i H'_{0} \tau / \hbar}\hat{X}_{m} e^{i H'_{0} \tau / \hbar}
\end{equation} with $H'_{0} = \hbar (\Delta_{XL}-\Delta_{P}) a^{\dagger} a + \hbar \Delta_{CL} c^{\dagger} c$ where we renormalize $\Delta_{XL}-\Delta_{P} \rightarrow \Delta_{XL}$. The component transformations become
\begin{align*}
e^{-i H'_{0} \tau / \hbar} a e^{i H'_{0} \tau / \hbar} &=  a^{\dagger} e^{-i\Delta_{XL}\tau} \\
e^{-i H'_{0} \tau / \hbar} a^{\dagger} e^{i H'_{0} \tau / \hbar} &=  a^{\dagger} e^{i\Delta_{XL}\tau} \\
e^{-i H'_{0} \tau / \hbar} c^{\dagger} a e^{i H'_{0} \tau / \hbar} &= c^{\dagger} e^{i\Delta_{CL}\tau} a e^{-i\Delta_{XL}\tau} \\
& = c^{\dagger} a e^{-i\Delta_{CX}\tau} \\
e^{-i H'_{0} \tau / \hbar} a^{\dagger} c e^{i H'_{0} \tau / \hbar} &= a^{\dagger} e^{i\Delta_{XL}\tau} c e^{-i\Delta_{CL}\tau} \\
& = a^{\dagger} c e^{i\Delta_{CX}\tau} \\
\end{align*} Without the coherent driving term, substitution gives
\begin{align*}
&G_{g}(\tau)[\hat{X}_{g},e^{-i H'_{S} \tau / \hbar}\hat{X}_{g} e^{i H'_{S} \tau / \hbar}\rho(t)] + H.c. \\ 
&\simeq \hbar^{2} g^{2} G_{g} (\tau) (c^{\dagger} a + a^{\dagger} c)(c^{\dagger} a e^{i\Delta_{CX}\tau} + a^{\dagger} c e^{-i\Delta_{CX}\tau}) \rho (t) \\
& - \hbar^{2} g^{2} G_{g} (\tau) (c^{\dagger} a e^{i\Delta_{CX}\tau} + a^{\dagger} c e^{-i\Delta_{CX}\tau}) \rho (t) (c^{\dagger} a + a^{\dagger} c) \\
&+ \hbar^{2} g^{2} G^{*}_{g} (\tau) \rho (t) (c^{\dagger} a e^{-i\Delta_{CX}\tau} + a^{\dagger} c e^{i\Delta_{CX}\tau}) (c^{\dagger} a + a^{\dagger} c) \\
&- \hbar^{2} g^{2} G^{*}_{g} (c^{\dagger} a + a^{\dagger} c) \rho (t) (c^{\dagger} a e^{-i\Delta_{CX}\tau} + a^{\dagger} c e^{i\Delta_{CX}\tau}) \\
& = \hbar^{2} g^{2} [G_{g}(\tau) e^{-i\Delta_{CX}\tau} \underbrace{a^{\dagger} c c^{\dagger} a \rho(t)}_{1} + G_{g}(\tau) e^{i\Delta_{CX}\tau} \underbrace{c^{\dagger} a a^{\dagger} c \rho(t)}_{5}] \\
&-\hbar^{2} g^{2} [G_{g}(\tau) e^{-i\Delta_{CX}\tau} \underbrace{c^{\dagger} a \rho(t) a^{\dagger} c}_{3} + G_{g}(\tau) e^{i\Delta_{CX}\tau} \underbrace{a^{\dagger} c \rho(t) c^{\dagger} a}_{7}] \\
&+ \hbar^{2} g^{2} [G^{*}_{g}(\tau) e^{-i\Delta_{CX}\tau} \underbrace{\rho(t) c^{\dagger} a a^{\dagger} c}_{6} + G^{*}_{g}(\tau) e^{i\Delta_{CX}\tau} \underbrace{\rho(t) a^{\dagger} c c^{\dagger} a}_{2}] \\
&- \hbar^{2} g^{2} [G^{*}_{g}(\tau) e^{-i \Delta_{CX} \tau} \underbrace{a^{\dagger} c \rho (t) c^{\dagger} a}_{8} + G^{*}_{g}(\tau) e^{i\Delta_{CX} \tau} \underbrace{c^{\dagger} a \rho (\tau) a^{\dagger} c}_{4}] \\
& = \hbar^{2} g^{2} \{ \underbrace{\Re[G_{g}(t) e^{-i \Delta_{CX} \tau}] a^{\dagger} c c^{\dagger} a \rho(t) + i \Im[G_{g}(\tau) e^{-i \Delta_{CX} \tau}] a^{\dagger} c c^{\dagger} a \rho(t)}_{1} \} \\
& + \hbar^{2} g^{2} \{ \underbrace{\Re[G_{g}(\tau) e^{-i \Delta_{CX} \tau}] \rho(t) a^{\dagger} c c^{\dagger} a - i \Im[G_{g}(\tau) e^{-i \Delta_{CX} \tau}] \rho(t) a^{\dagger} c c^{\dagger} a}_{2} \} \\
& - \hbar^{2} g^{2} \{ \underbrace{G_{g}(\tau) e^{-i \Delta_{CX} \tau} c^{\dagger} a \rho(t) a^{\dagger} c}_{3} + \underbrace{G^{*}_{g}(\tau) e^{i \Delta_{CX} \tau} c^{\dagger} a \rho(t) a^{\dagger} c}_{4} \} \\
& + \hbar^{2} g^{2} \{ \underbrace{\Re[G_{g}(\tau) e^{i \Delta_{CX} \tau}] c^{\dagger} a a^{\dagger} c \rho(t) + i \Im[G_{g}(\tau) e^{i \Delta_{CX} \tau}] c^{\dagger} a a^{\dagger} c \rho(t)}_{5} \} \\
& + \hbar^{2} g^{2} \{ \underbrace{\Re[G_{g}(\tau) e^{i \Delta_{CX} \tau}] \rho(t) c^{\dagger} a a^{\dagger} c - i \Im[G_{g}(\tau) e^{i \Delta_{CX} \tau}] \rho(t) c^{\dagger} a a^{\dagger} c}_{6} \} \\
& - \hbar^{2} g^{2} \{ \underbrace{G_{g}(\tau) e^{i \Delta_{CX} \tau} a^{\dagger} c \rho(t) c^{\dagger} a}_{7} + \underbrace{G^{*}_{g}(\tau) e^{-i \Delta_{CX} \tau} a^{\dagger} c \rho (t) c^{\dagger} a}_{8} \} \\
\end{align*}
\begin{align*}
&G_{g}(\tau)[\hat{X}_{g},e^{-i H'_{S} \tau / \hbar}\hat{X}_{g} e^{i H'_{S} \tau / \hbar}\rho(t)] + H.c. \\ 
& \simeq \hbar^{2} g^{2} \Re[G_{g}(\tau) e^{-i \Delta_{CX} \tau}] ( \underbrace{a^{\dagger} c c^{\dagger} a \rho(t)}_{\Re[1]} + \underbrace{\rho(t) a^{\dagger} c c^{\dagger} a}_{\Re[2]} - \underbrace{2 c^{\dagger} a \rho(t) a^{\dagger} c}_{3+4}) \\
& + \hbar^{2} g^{2} \Re[G_{g}(\tau) e^{i \Delta_{CX} \tau}] ( \underbrace{c^{\dagger} a a^{\dagger} c \rho(t)}_{\Re[5]} + \underbrace{\rho(t) c^{\dagger} a a^{\dagger} c}_{\Re[6]} - \underbrace{2 a^{\dagger} c \rho(t) c^{\dagger} a}_{7+8}) \\
& + i \hbar^{2} g^{2} \Im[G_{g}(\tau) e^{-i \Delta_{CX} \tau}] ( \underbrace{a^{\dagger} c c^{\dagger} a \rho(t)}_{\Im[1]} - \underbrace{\rho(t) a^{\dagger} c c^{\dagger} a}_{\Im[2]}) \\
& + i \hbar^{2} g^{2} \Im[G_{g}(\tau) e^{i \Delta_{CX} \tau}] ( \underbrace{c^{\dagger} a a^{\dagger} c \rho(t)}_{\Im[5]} - \underbrace{\rho(t) c^{\dagger} a a^{\dagger} c}_{\Im[6]} ) 
\end{align*} We lose eight terms in the first equality by cancellation with terms from the integrand containing $X_{u}$ due to the imaginary $i$ giving an overall minus sign to the manipulations. By inspection of the last equality all $c^{\dagger}$ are replaced by $-c^{\dagger}$ for the integrand with $X_{u}(\tau)$ but the overall minus sign negates this. We can then sum over $m=g,u$.
\begin{align*}
g^{2} \int_{0}^{\infty} d\tau \sum_{m=g,u} & \{ \Re[G_{m}(\tau) e^{-i \Delta_{CX} \tau}] ( a^{\dagger} c c^{\dagger} a \rho(t) + \rho(t) a^{\dagger} c c^{\dagger} a - 2 c^{\dagger} a \rho(t) a^{\dagger} c) \\
& + \Re[G_{m}(\tau) e^{i \Delta_{CX} \tau}] ( c^{\dagger} a a^{\dagger} c \rho(t) + \rho(t) c^{\dagger} a a^{\dagger} c - 2 a^{\dagger} c \rho(t) c^{\dagger} a) \\
& + i \Im[G_{m}(\tau) e^{-i \Delta_{CX} \tau}] ( a^{\dagger} c c^{\dagger} a \rho(t) - \rho(t) a^{\dagger} c c^{\dagger} a ) \\
& + i \Im[G_{m}(\tau) e^{i \Delta_{CX} \tau}] ( c^{\dagger} a a^{\dagger} c \rho(t) - \rho(t) c^{\dagger} a a^{\dagger} c ) \}
\end{align*} Recall $\sum_{m=g,u} G_{m}(\tau) = \langle B \rangle^{2} (e^{\phi(\tau)} - 1)$ and the dissipator is defined $\mathcal{L}(\xi) = \xi\rho\xi^{\dagger}-\frac{1}{2}\xi^{\dagger}\xi\rho-\frac{1}{2}\rho\xi^{\dagger}\xi$ with Lindbladian operators $\xi$. The resulting term becomes
\begin{equation}
i [\Delta^{a^{\dagger} c}_{ph} a^{\dagger} c c^{\dagger} a + \Delta^{c^{\dagger} a}_{ph} c^{\dagger} a a^{\dagger} c,\rho(t)] + \frac{\Gamma^{a^{\dagger} c}_{ph}}{2} \mathcal{L}(a^{\dagger} c) + \frac{\Gamma^{c^{\dagger} a}_{ph}}{2} \mathcal{L}(c^{\dagger} a)
\end{equation} The frequency shifts are given
\begin{equation}
\Delta^{a^{\dagger}c/c^{\dagger}a}_{ph} = \langle B \rangle^{2} g^{2} \mathrm{Im}\left [ \int_{0}^{\infty} d\tau e^{\pm i \Delta_{CX} \tau} (e^{\phi(\tau)}-1) \right ]
\end{equation} The scattering rates are given
\begin{equation}
\Gamma^{a^{\dagger}c/c^{\dagger}a}_{ph} = 2 \langle B \rangle^{2} g^{2} \mathrm{Re}\left [ \int_{0}^{\infty} d\tau e^{\pm i \Delta_{CX} \tau} (e^{\phi(\tau)}-1) \right ]
\end{equation} If we ignore cross-terms in the integrand commutator between the coherent exciton-cavity interaction $g$ and coherent exciton drive $\eta_{X}$ we can avoid a rederivation by observing we can remove all $c$ and $c^{\dagger}$, replacing $g \rightarrow \eta_{X}$, and replacing $\Delta_{CX} \rightarrow \Delta_{XL}$.  The resulting term becomes
\begin{equation}
i [\Delta^{a^{\dagger}}_{ph} a^{\dagger} a + \Delta^{a}_{ph} a a^{\dagger},\rho(t)] + \frac{\Gamma^{a^{\dagger}}_{ph}}{2} \mathcal{L}(a^{\dagger}) + \frac{\Gamma^{a}_{ph}}{2} \mathcal{L}(a)
\end{equation} The frequency shifts are given
\begin{equation}
\Delta^{a^{\dagger}/a}_{ph} = \langle B \rangle^{2} \eta_{X}^{2} \mathrm{Im}\left [ \int_{0}^{\infty} d\tau e^{\pm i \Delta_{XL} \tau} (e^{\phi(\tau)}-1) \right ]
\end{equation} The scattering rates are given
\begin{equation}
\Gamma^{a^{\dagger}/a}_{ph} = 2 \langle B \rangle^{2} \eta_{X}^{2} \mathrm{Re}\left [ \int_{0}^{\infty} d\tau e^{\pm i \Delta_{XL} \tau} (e^{\phi(\tau)}-1) \right ]
\end{equation}

For the purposes of simulation and fitting to the data we choose to ignore the Stark shifts $\Delta^{a^{\dagger}c/c^{\dagger}a}_{ph}$ and $\Delta^{a^{\dagger}/a}_{ph}$.

\section{Photoluminescence}

\begin{figure}
	\centering
	\includegraphics[width=0.7\linewidth]{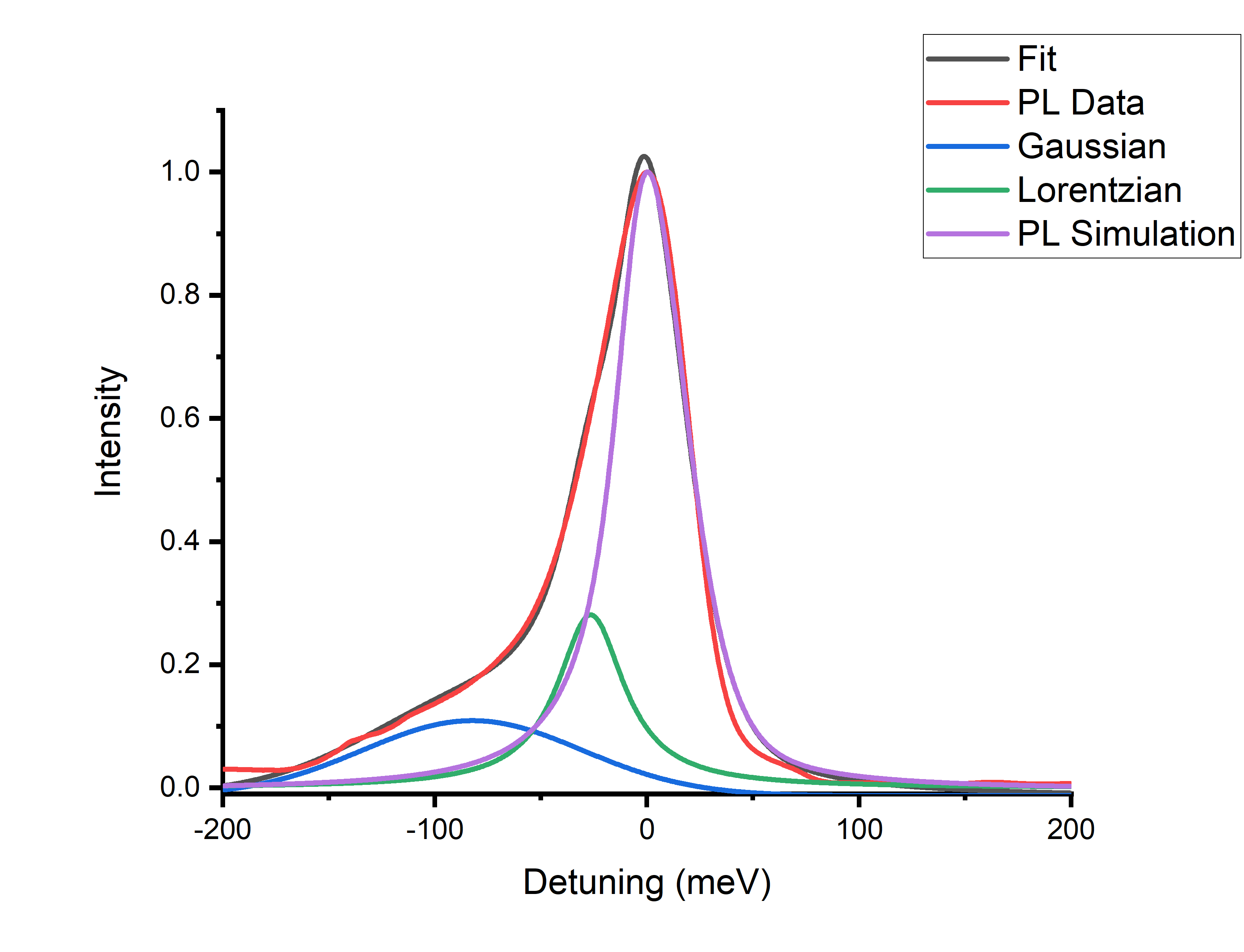}
	\caption{The extracted fit parameters for equation \ref{PLfit} are: $y_{0} = -0.01262$, $A_{D} = 15.82$, $\Gamma_{D} = 122.0$ \si{\milli\electronvolt}, $\Delta_{D} = -82.31$ \si{\milli\electronvolt}, $A_{T} = 17.02$, $\Gamma_{T} = 38.56$ \si{\milli\electronvolt}, $\Delta_{T} = -26.55$ \si{\milli\electronvolt}, $f = 0.9038$, $\alpha_{p} = 0.018$ \si{\pico \second}$^2$, and $\omega_{b} = 6.7$ \si{\milli\electronvolt}.}
	\label{fig:plfit}
\end{figure}

Our experiments involve above-band excitation which phenomenologically amounts to an incoherent drive of the neutral exciton. Our presented simplified model requires a coherent drive for phonon-mediated processes as seen by the $\eta_{X}$ in the $\Gamma^{a^{\dagger}/a}_{ph}$ scattering rates. If we represent the incoherent excitation as a coherent field with a random phase \cite{agarwal_additional_1991,carmichael_classical_1991,tian_incoherent_1992} the Markovian approximation made for the TCL master equation allows us to consider a steady state population of neutral excitons by effectively integrating out the transient dynamics.

The steady-state exciton population \cite{weiler_phonon-assisted_2012} without cavity-coupling can be found for comparison to photoluminescent (PL) measurements
\begin{equation}\label{exciton_population}
\bar{N}_{X} = \frac{1}{2}\left [ 1 + \frac{\Gamma^{\sigma^{+}}_{ph} - \Gamma^{\sigma^{-}}_{ph} - \gamma}{\Gamma^{\sigma^{+}}_{ph} + \Gamma^{\sigma^{-}}_{ph} + \gamma + \frac{4 \eta_{x}^{2} \langle B \rangle^{2} \Gamma_{pol}}{\Gamma^{2}_{pol} + \Delta_{XL}^{2}}} \right ]
\end{equation}
with $\Gamma_{pol} = \frac{1}{2}(\Gamma^{\sigma^{+}}_{ph} + \Gamma^{\sigma^{-}}_{ph} + \gamma)$. For numerical fitting of the data to this model we use $\eta_{X} = 0.01 \gamma$. The coefficient choice does not significantly affect the lineshape of $\bar{N}_x$. Fixed parameters in this model are temperature ($T$) and the exciton decay rate ($\gamma$) extracted from the linewidth of the exciton resonance. $\bar{N}_{X}$ reduces to a Lorentzian with full width at half maximum (FWHM) equal to $\gamma$. Free parameters of the model are the exciton-phonon coupling strength ($\alpha_{p}$) and cutoff frequency ($\omega_{b}$).

The secondary sidebands assumed to be defects or trions \cite{chow_defect-induced_2015,sidler_fermi_2017} are fit using a Gaussian and Lorentzian function, respectively. The defect Gaussian function incorporates an inhomogenous broadening mechanism due to local perturbations. The trion Lorentzian function assumes only homogeneous broadening \cite{ajayi_approaching_2017}. The total PL spectrum is fit using SciPy's $\boldsymbol{curve\_fit}$ function for the Gaussian and Lorentzian terms, and a brute force grid search for $\alpha_{p}$ and $\omega_{b}$.
\begin{equation}\label{PLfit}
S_{PL}(\Delta_{XL}) = y_{0} + \frac{A_{D} e^{\frac{-4 \ln{2} (\Delta_{XL} - \Delta_{D})^{2}}{\Gamma_{D}^{2}}}}{\Gamma_{D} \sqrt{\frac{\pi}{4 \ln{2}}}} + \frac{2 A_{T}}{\pi} \frac{\Gamma_{T}}{4 (\Delta_{XL} - \Delta_{T})^{2} + \Gamma_{T}^{2}} + f \bar{N}_{X}(\Delta_{XL})
\end{equation}
The defect ($\Delta_{D} = -82.31$ \si{\milli\electronvolt}) and trion ($\Delta_{T} = -26.55$ \si{\milli\electronvolt}) peak detunings with respect to the neutral exciton are consistent with previous results in the literature ($\Delta_{D} \approx -101$ \si{\milli\electronvolt}, $\Delta_{T} \approx -28$ \si{\milli\electronvolt}) \cite{chow_defect-induced_2015,li_revealing_2018}. 

Although there are excellent papers on the subject of phonon-mediated interactions in monolayer materials \cite{moody_intrinsic_2015,dey_optical_2016,christiansen_phonon_2017,shree_observation_2018}, there is no clear comparison for $\alpha_{p}$ and $\omega_{b}$. As a sanity check, $\alpha_{p}$ for a spherically confining potential is known to be $\frac{(D_{c}-D_{v})^{2}}{4 \pi^{2} \rho s^{5}}$ \cite{calarco_spin-based_2003,nazir_photon_2008,xue_detuning_2008}. $D_{c}$ ($D_{v}$) are the deformation potential constants for the conduction (valence) band, $\rho$ is the bulk material density, and $s$ is the sound velocity. Using the bulk material parameters for $\text{WSe}_{2}$ with $|D_{c} - D_{v}| = 5.4$ \si{\electronvolt} \cite{schmidt_reversible_2016}, $\rho = 9.32$ \si{\gram \centi\meter}$^{-3}$ \cite{agarwal_growth_1979}, and $s = 4000$ \si{\meter\second}$^{-1}$ \cite{jin_intrinsic_2014} we find $\alpha_{p} = 0.019$ \si{\milli\electronvolt}, which is similar to our extracted result of $\alpha_{p} = 0.018$ \si{\milli\electronvolt}. The phonon cutoff energy $\omega_{b} = \frac{s}{d} = 6.7$ \si{\milli \electronvolt} corresponds to a $d=3$ \si{\nano \meter} in-plane localization. This length scale is not unreasonable for an estimate of the silicon nitride surface roughness. There may also be some correlation between the localization and defect emission.

\section{Total Spectrum}

The cavity-coupled PL spectrum is found from the Fourier transform of the correlation function
\begin{equation*}
S(\omega) = \int_{-\infty}^{\infty} \lim_{t \rightarrow \infty} \langle c^{\dagger}(t+\tau)c(t)\rangle d\tau
\end{equation*} which is provided as a function in the QuTiP Python library \cite{johansson_qutip:_2012,johansson_qutip_2013}. Although it is tempting to simultaneously include all cavity resonances in the system Hamiltonian, with the cavity bosonic operator truncated at $N=10$ to ensure convergence we quickly approach a Hilbert space dimension that is classically intractable to simulate. Instead, we recognize each cavity mode of the ring resonator can be treated independently because the spatial overlap of different modes is approximately equal to zero. Similarly, we only consider cavity modes which have the same spatial wavefunction as the collective excitation of the excitonic resonance \cite{verger_polariton_2006,munoz-matutano_emergence_2019}. The total spectrum $S_{Total}$ is then a linear superposition of all cavity resonances with different detuning $\Delta_{CX}$ coupled to the monolayer PL and background PL
\begin{align*}
S_{Total}(\omega) &= f_{1} S_{PL}(\omega) + f_{2} \sum_{\Delta_{CX}} S_{\Delta_{CX}}(\omega) \\ &+ f_{3} \frac{A_{D} e^{\frac{-4 \ln{2} (\omega - \Delta_{D})^{2}}{\Gamma_{D}^{2}}}}{\Gamma_{D} \sqrt{\frac{\pi}{4 \ln{2}}}} \sum_{\Delta_{CX}} \frac{2 A_{DC}}{\pi} \frac{\Gamma_{DC}}{4 (\omega - \Delta_{CX})^{2} + \Gamma_{DC}^{2}} \\ &+ f_{4} \frac{2 A_{T}}{\pi} \frac{\Gamma_{T}}{4 (\omega - \Delta_{T})^{2} + \Gamma_{T}^{2}} \sum_{\Delta_{CX}} \frac{2 A_{TC}}{\pi} \frac{\Gamma_{TC}}{4 (\omega - \Delta_{CX})^{2} + \Gamma_{TC}^{2}}
\end{align*} The free parameters for fitting the model to the cavity-coupled PL data are $f_{1-4}$ as the relative intensity of the background PL, cavity-coupled neutral exciton PL, cavity-coupled defect PL and cavity-coupled trion PL and the exciton-cavity coupling $g$. The relative intensity in our experiments is found to be $f_{1} = 0.53, f_{2} = 0.44, f_{3} = 2.18$, and $f_{4} = 1.49$. We chose a single cavity parameter for defect and trion emission to reduce overfitting. For the cavity-coupled defect PL $A_{DC} = 2.24$ and $\Gamma_{DC} = 2.33$ \si{\milli \electronvolt}. For the cavity-coupled trion PL $A_{TC} = 1.42$ and $\Gamma_{TC} = 2.29$ \si{\milli \electronvolt}. 

\bibliography{references}
\bibliographystyle{unsrtnat}
	
\end{document}